# Hidden spin-texture at topological domain walls drive exchange bias in a Weyl semimetal

A. Noah[1], F. Toric[1], T.D. Feld[1], G. Zissman[1], A. Gutfreund[1], D. Tsruya[1], T. R. Devidas[1], H. Alpern[1], H. Steinberg[1], M.E. Huber[2], J.G. Analytis[3,4], S. Gazit[1,5], E. Lachman[3,4] and Y. Anahory[1]

[1]The Racah Institute of Physics, The Hebrew University, Jerusalem, 91904, Israel
[2]Departments of Physics and Electrical Engineering, University of Colorado Denver, Denver, CO 80217, USA
[3]Department of Physics, University of California, Berkeley, CA 94720, USA
[4]Materials Sciences Division, Lawrence Berkeley National Laboratory, Berkeley, CA, 94720, USA
[5]The Fritz Haber Research Center for Molecular Dynamics, The Hebrew University of Jerusalem, Jerusalem 91904, Israel

**Abstract:** Exchange bias is a phenomenon critical to solid-state technologies that require spin valves or non-volatile magnetic memory. The phenomenon is usually studied in the context of magnetic interfaces between antiferromagnets and ferromagnets, where the exchange field of the former acts as a means to pin the polarization of the latter. In the present study, we report an unusual instance of this phenomenon in the topological Weyl semimetal $Co_3Sn_2S_2$, where the magnetic interfaces associated with domain walls suffice to bias the entire ferromagnetic bulk. Remarkably, our data suggests the presence of a hidden order parameter whose behavior can be independently tuned by applied magnetic fields. For micron-size samples, the domain walls are absent, and the exchange bias vanishes, suggesting the boundaries are a source of pinned uncompensated moment arising from the hidden order. The novelty of this mechanism suggests exciting opportunities lie ahead for the application of topological materials in spintronic technologies.

## Introduction

The phenomenon of exchange bias (EB) was first realized in Co/CoO$_x$ heterostructure particles[1]. The magnetic hysteresis loop of this heterostructure was centered around a field $H_{EB} = -(H_c^- + H_c^+)/2 \neq 0$, where $H_c^+$ and $H_c^-$ represent the coercive fields for the positive and negative fields, respectively. The observed shift in the magnetic hysteresis loop was attributed to the pinning of the ferromagnetic (FM) Co moments by the exchange interactions with the anti-ferromagnetic (AFM) CoO$_x$. Since then, EB was observed in other heterostructures of different FM and AFM materials[2] including spin glasses[3], superparamagnets[4], and ferrimagnets[5]. Moreover, some realizations of EB can be produced without heterostructures[6,7], and some do not require activation by cooling in a magnetic field and create spontaneous EB that is isothermally set at low temperatures[4,8]. This abundance of systems presenting EB have produced a plethora of suggested mechanisms to explain this phenomenon. The main component of these mechanisms lies in a source of uncompensated spins that induce a preferred direction.

Recently, EB was observed in a pure bulk $Co_3Sn_2S_2$ single crystals[9], with no substrate effects[6] or material doping[7], invoking the need for a different mechanism for the EB they present. $Co_3Sn_2S_2$ was recently discovered as a magnetic Weyl semimetal, where the breaking of time-reversal symmetry due to ferromagnetism satisfy the requirement of a broken symmetry to create the topological band structure[10–18]. Of particular interest is the non-trivial interplay between strong magnetic correlations and the emergent topological band structure of Weyl fermions in such materials[19–21], which is not present in their inversion-symmetry-breaking counterparts. Specifically, the magnetic moments in $Co_3Sn_2S_2$ reside on the Co sites and are arranged in a layered kagome lattice. Experimentally, below 175 K, a dominant FM phase with an easy axis oriented out of the kagome plane is observed, although this phase may coexist with an in-plane (IP) antiferromagnetic phase[9,20]. Below this temperature, the electronic dispersion has been shown to host Weyl fermions[16,22] associated with a giant anomalous Hall effect[12,13].

Even at the level of basic magnetic properties, experimental observations display puzzling inconsistencies. Explicitly, the reported values of magnetic saturation field, $H_s$, range from 86 mT to 900 mT in the bulk[9,12,13,20,23,24], and may reach values of a few Tesla in micron-size samples[25–28]. Such large variations are surprising since $H_s$ is a material property determined by the crystalline structure and electronic orbitals, and the large variability is unlikely to be due to a simple geometrical factor or sample quality. Presently, these unusual magnetic properties including the presence of EB and their underlying physics are far from understood.

In this work, we address these questions by using a combination of transport measurements and magnetic imaging to visualize the magnetic domains and elucidate the mechanism underlying the EB in $Co_3Sn_2S_2$. We reveal a new kind of EB, that can be set and changed isothermally at low temperatures. We attribute this to a hidden IP magnetic order that is likely related to the domain walls. These findings suggest a promising future for the application of magnetic Weyl semimetals in memory and spintronic technologies.

**Results**

$Co_3Sn_2S_2$ crystals were grown by the self-flux method commonly used for these crystals[9,29,30]. For electronics transport measurements, four aluminum wires were glued to a 80 μm-thick sample using silver epoxy, as shown in Fig. 1a (see Methods and Supplementary Note 5). The sample was zero-field cooled (ZFC) down to 4.2 K, where the sample displays FM characteristics [20]. A 2 mA current was applied along the $x$ axis while the voltage $V_y$ was measured along the $y$ axis (see Fig. 1a). We report the Hall resistance $R_{xy} = V_y/I_x$ as a function of the out-of-plane (OOP) magnetic field $\mu_0 H_z$ under distinct magnetic field protocols up to 6 Tesla.

As an example, Figure 1b (black curve) presents $R_{xy}$ measurements starting from ZFC and ramping $H_z$ at a rate of 4 mT/s to +1 T. Subsequently, we carried out a full −1 T to +1 T magnetic field loop. During the initial field ramp, $R_{xy}$ grows smoothly, starting from zero and reaching saturation at 95 mT (see Fig. 3b blue curve). The following $R_{xy}$ curve exhibits sharp jumps associated with two distinct coercive fields, $H_c^+ = 140$ mT and $H_c^- = -240$ mT, for positive and negative $H_z$, respectively (Fig. 1b, black curve). The asymmetry between the coercive fields yields $H_{EB} = 50$ mT. We note that these $H_c^\pm$ values are reproducible within our experimental resolution (better than 5 mT), as long as the field is swept symmetrically from +1 T to −1 T ($\mu_0|H_{max}^\pm|$ = 1 T). As previously[9] reported for ZFC conditions, the lower coercive field is determined by the direction of the first field excursion. Here the positive direction was chosen initially, and we obtain $|H_c^+| < |H_c^-|$, and thus $H_{EB} > 0$.

Our results differ from those in previous reports, which did not observe a finite $H_{EB}$ when performing a magnetization loop with a much larger $\mu_0 H_{max}^\pm \sim 9$ T. It is therefore interesting to explore the evolution of the hysteresis loop at intermediate fields, and for asymmetric field-sweep protocols. With that goal in mind, we apply such a protocol with $\mu_0 H_{max}^- = -1.5, -2, -3, -6$ T while fixing $\mu_0 H_{max}^+ = 1$ T (Fig. 1b blue curves). Interestingly, we find that $H_c^+$ values increase by more than a factor of 5, ranging from 140 mT ($\mu_0 H_{max}^- = -1$ T) to 736 mT ($\mu_0 H_{max}^- = -6$ T). In contrast, $H_c^-$ varies only slightly from the -240 mT observed in the $\pm 1$ T symmetric sweep to -140 mT for all the other sweeps. Consequently, over these protocols, $H_{EB}$ changes both sign and magnitude from $H_{EB} = 50$ mT after ZFC, to -300 mT after $\mu_0 H_{max}^- = -6$ T. Different values of $|H_{max}^\pm|$ yield different values of $H_c^\mp$ (Fig. 1b inset). Curiously, $H_c^+$ remains constant for distinct values of $H_{max}^-$ suggesting that $H_c$ increases in discrete steps. The observed sign reversibility in $H_{EB}$ implies that it can be set to be vanishingly small, as demonstrated in Supplementary Note 2.

It is essential to contrast the above result with those from minor loop protocols used in conventional ferromagnets, where distinct $H_c$ values can be reached by applying $|H_{max}| < H_s$. Physically, this phenomenon can be attributed to residual magnetic domains that are anti-parallel with the external field[31,32]. The present case is different, since $|H_{max}| > H_s$ in all the protocols described above. Our measured $H_s$ is consistent with previous reports where $H_s < 900$ mT[9,12,13,20,23,24].

The non-trivial evolution of $H_c^\pm$ as a function of $|H_{max}|$ beyond $H_s$ indicates that $H_c^\pm$ does **not** depend solely on OOP magnetization. Notably, previous reports could not rule out a slight canting of magnetic moments below 125K, even though the magnetization was predominantly oriented OOP[12,20]. Moreover, theoretical calculations predict that even slight canting might dramatically affect the Weyl nodes[33] and associated magnetic textures[19]. Motivated by these theoretical predictions, it is also interesting to study the effect of an IP field, $H_{IP}$, on $H_c^\pm$.

To reveal the influence of $H_{IP}$ on $H_c^\pm$, we conduct IP field excursions $H_{IP}^{ex} = \pm 2$ T, followed by the $\mu_0|H_{max}^\pm|$ = 1 T loop protocol at $H_{IP} = 0$. Considering that the IP saturation field is 23 T[12,24], this should naively tilt the spin by $\sim 8°$. We start by measuring $R_{xy}$ after ZFC, before applying any IP field. Under these conditions, $R_{xy}$ shows the expected $H_c^+ = 128$ mT and $H_c^- = -241$ mT with $H_{EB} = 56$ mT (black line, Fig. 1c). An IP field excursion reaching $\mu_0 H_{IP}^{ex} = 2$ T, reveals a small increase in both $H_c^+ =139$ mT, $H_c^- = -275$ mT and $H_{EB} = 68$ mT (blue line) with respect to the ZFC measurements. On the other hand, an IP field excursion in the opposite direction $\mu_0 H_{IP}^{ex} = -2$ T, results in a change of sign in $H_{EB}$, with $H_c^+ = 244$ mT, $H_c^- = -128$ mT, and $H_{EB} = -58$ mT (red line). This experimental observation indicates that even a small degree of canting has a dramatic effect on $H_{EB}$. We emphasize that the material retains a memory of the $H_{EB}$ magnitude and sign, even after the IP field is set to zero. We also note that no hysteresis behavior is observed for the IP magnetization as a function of the IP field [12,24]. Other measurements in the presence of a finite $H_{IP}$ are presented in supplementary note 3.

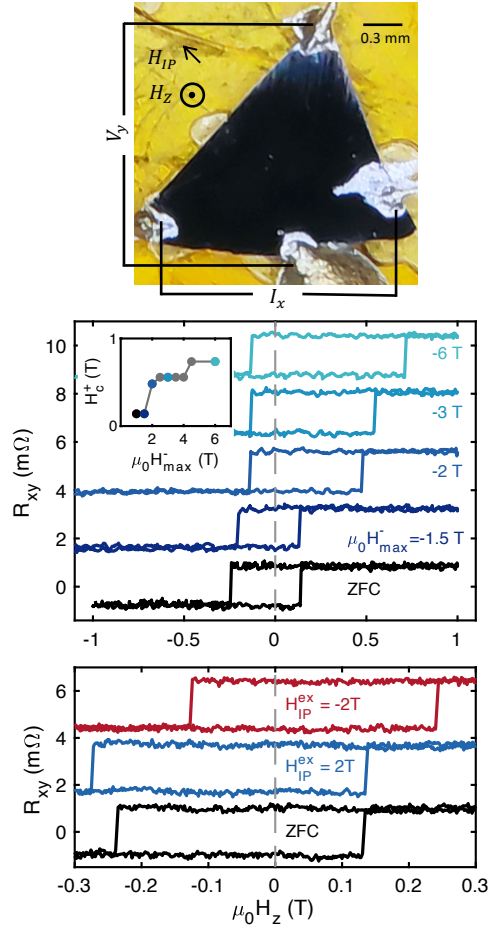

**Fig. 1 Controlling the exchange bias of Co₃Sn₂S₂ single crystal sample by applying out-of-plane and in-plane field protocols at 4.2 K.** (**a**) Optical image of the sample and a schematic depicting paths for current $I_x$ and voltage probe $V_y$. (**b**) Hall resistance $R_{xy} = V_y/I_x$ as a function of the out-of-plane field $\mu_0 H_z$. After ZFC the field is swept between $\pm 1$ T resulting in a rectangular $R_{xy}$ with $H_{EB} = 50$ mT (black). controlling the positive coercive field $H_c^+$ by performing a larger field excursion in the negative direction $\mu_0 H_{max}^- = -1.5, -2, -3, -6$ T while keeping $\mu_0 H_{max}^+ = 1$ T resulting in $H_{EB} = 35, -170, -200, -300$ mT, respectively (dark blue to light blue). Inset: $H_c^+$ as a function of $\mu_0 H_{max}^-$, with colored dots corresponding to the measurements shown in **b**. (**c**) Each measurement is performed at $H_{IP} = 0$ after ZFC (black), after an excursion at $H_{IP}^{ex} = 2$ T resulting in $H_c^+ = 139$ mT and $H_c^- = -275$ mT, $H_{EB} = 68$ mT (blue) and at $\mu_0 H_{IP}^{ex} = -2$ T yielding $H_c^+ = 244$ mT, $H_c^- = -128$ mT and $H_{EB} = -58$ mT (red). Three loops performed for each measurement gave similar results, and one representative result is shown for clarity. The curves for different protocols are shifted vertically for clarity.

In order to further explore the possibility that the $H_{EB}$ information is not encoded in the OOP magnetization, we now attempt to demagnetize the sample, using a minor loop protocol. Explicitly, we carry out one minor loop sweep with $\mu_0|H_{max}^{\pm}| \sim H_c^{\pm}$ (Fig. 2a Red curve number 1 and 2). The results reveal a vanishingly small $R_{xy}$ (number 3), which naively indicates that the sample is demagnetized. A priori, one would expect a demagnetized sample to lose all $H_{EB}$ information and return to the initial random ZFC conditions. To test this expectation, we carry out an OOP field excursion oppositely aligned with the initial ZFC excursion. Remarkably, we find precisely the same hysteresis loop as after the ZFC excursion (Fig. 2b blue curve). This finding suggests that even though the sample appears demagnetized in transport, it retains the information regarding the direction of the first field excursion after ZFC. More broadly, this provides experimental evidence that $H_{EB}$ information is stored in degrees of freedom other than the global OOP magnetization.

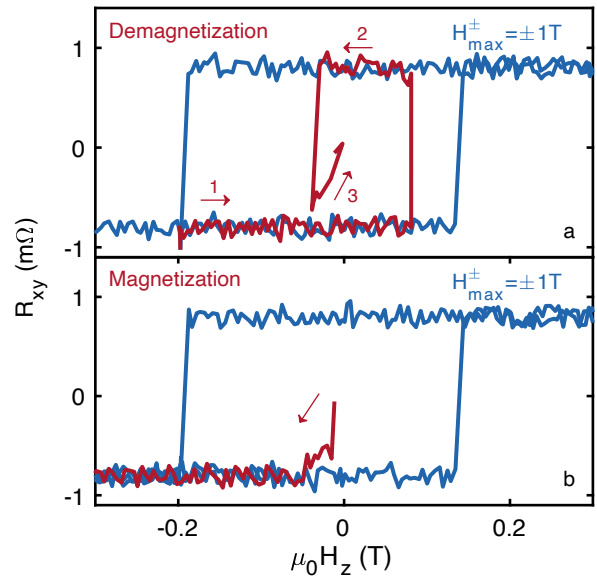

**Fig. 2 Demagnetization and magnetization (a)** Demagnetization protocol, the field swept from -1T to 1T to initialize the sample and from +1 until $H_c^-$ is reached (blue). Sweeping the field between the distinct $H_c^{\pm}$, after three repetitions produces a vanishingly small $R_{xy}$ (red). **(b)** Magnetization of the sample. The field is first swept to -1T (red) and then a +/- 1T loop is performed (blue).

To gain a better insight into the magnetic structure, we conduct local magnetic field imaging $B_z(x,y)$ using a scanning SQUID that resides at the end of a sharp tip (Fig. 3a). The SQUID-on-tip (SOT) provides high spatial resolution for magnetic imaging[34,35] reaching single-spin sensitivity[36,37]. Images are taken at 4.2 K with SQUIDs having a diameter ranging from 110 to 130 nm (see Methods and Supplementary Note 1).

Fig. 3c presents the results of imaging the ZFC sample with magnetic features of a few microns yielding a magnetic contrast of ~ 50 mT. With increasing field, domains parallel to the field grow at the expense of the anti-parallel domains (Fig. 3 c-f and Movie 1). Above the saturation field, 95 mT, starting from ZFC conditions, the magnetic contrast drops below 1.5 mT. The local magnetic structure does not appear to further evolve on this scale, which is in agreement with transport and global magnetization measurements[9,12,13,20,23,24].

As shown previously, while the ZFC and the demagnetized state both exhibit a vanishing $R_{xy}$, there is a crucial difference between the samples in that the latter retains the $H_{EB}$ memory, while the former does not. It is therefore interesting to observe whether this difference is also apparent in the local magnetic structure. To examine this, we compare the respective magnetic structures of the two states (Fig. 3 g-j and movie 2). The results indicate that, unlike the ZFC state, the demagnetized state exhibits a stripe pattern with a typical width of 10 μm, and a length that extends beyond our field of view (45 μm). The memory of a finite exchange bias $H_{EB}$ is retained only by the larger domain.

Similar magnetic domain structures are observed when the magnetization reverses (Fig 3 k-n and movie 3). Notably, the appearance of these features always coincides precisely with a sign change in $R_{xy}$, indicating that our microscopic images (45 X 45 μm$^2$) are representative of the much larger sample (mm size). We note that this transitory state is observed only for loops where $|H^{\pm}_{max}|$ is relatively low $|H^{\pm}_{max}| < 300$ mT, and is only visible for less than 30 mT beyond the field at which $R_{xy}$ changes sign. For protocols with a larger $|H^{\pm}_{max}|$, no transitory state was observed and instead an abrupt change in local magnetic field was recorded by the SOT between two field steps (smaller than 5 mT) coinciding with the change of sign of $R_{xy}$. This indicates that the magnetization reversal in such case occurs abruptly throughout the sample with no evolution of the magnetic landscape observed on either side of the magnetic transition.

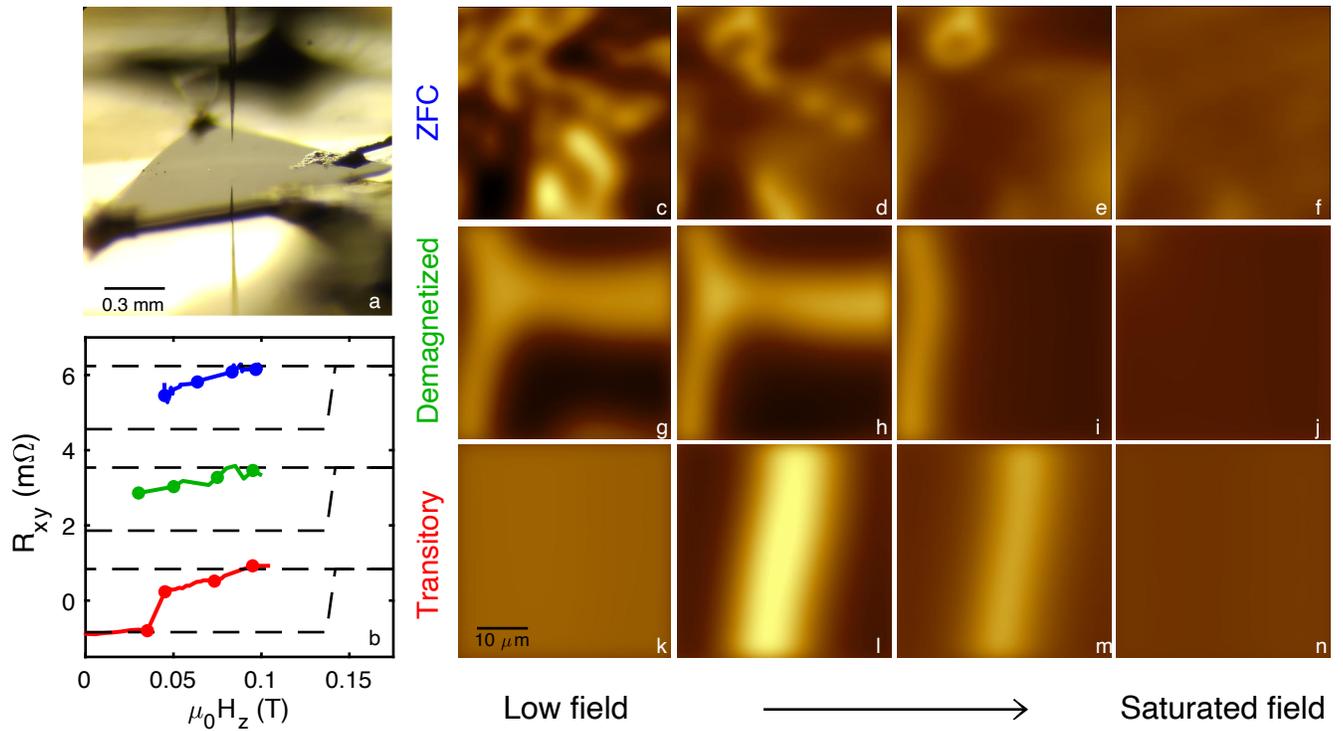

**Fig. 3 Scanning SOT microscopy images of Co$_3$Sn$_2$S$_2$ bulk at 4.2 K.** (**a**) Optical image of the SOT pointing down, the sample, and the reflection of the SOT on the sample. (**b**) $R_{xy}$ as a function of applied magnetic field for different phase of the sample. ZFC to saturated field (blue), demagnetized state to saturated field (green) and magnetization reversal with transitory state (red). $R_{xy}$ corresponding to a loop with $\mu_0|H^{\pm}_{max}| = 1$ T is shown for comparison (black dashed). The fields at which the images were taken are marked with dots. The curves for different protocols are shifted vertically for clarity. (**c-n**) Sequence of magnetic images of different state of the sample at distinct values of applied out-of-plane field $\mu_0 H_z$. (**c-f**). Evolution from ZFC to the saturation field. $\mu_0 H_z = 45$ mT **c**, $\mu_0 H_z = 65$ mT **d**, $\mu_0 H_z = 85$ mT **e**, $\mu_0 H_z = 95$ mT **f**. (**g-j**) Evolution of demagnetization to the saturation field. $\mu_0 H_z = 30$ mT **g**, $\mu_0 H_z = 50$ mT **h**, $\mu_0 H_z = 75$ mT **i**, $\mu_0 H_z = 95$ mT **j**. (**k-n**) Magnetization reversal with transitory state. $\mu_0 H_z = 35$ mT **k**, $\mu_0 H_z = 45$ mT **l**, $\mu_0 H_z = 75$ mT **m**, $\mu_0 H_z = 95$ mT **n**. All images are 45x45 μm$^2$, pixel size 480 nm **c-f** and 980 nm **g-n**, acquisition time 8 min/image **c-f** and 5.8 min/image **g-n**. The bright to dark color scale represents 50 mT and is the same for all images. See Supplementary Movies 1-3 corresponding to images **c-f**, **g-j**, **k-n**, respectively.

From above observations, we can conjecture that there is a minimal length scale of the domains required for the appearance of EB. This hypothesis can be tested by repeating the same field sweep protocols on samples with lateral dimensions of a few tens of microns, which is comparable to the size of the domains. For such samples, we expect to see a single, or at most a few, magnetic domains. We use a Focused Ion Beam (FIB), to cut a slab of area of 60 x 30 µm² from the bulk crystal with thicknesses ranging from 1 µm to 42 µm and the $c$ axis pointing out of the plane (Fig. 4a). Platinum is then used as a contact to lithographically defined 50 nm-thick Nb tracks to the sample. Chemical analysis reveals that the sample maintained stoichiometric ratio of its constituent elements except near the surface where C and Ga were detected (Supplementary Note 6). The FIB sample is ZFC only down to 10 K in order to avoid the complications arising from the superconductivity in Nb ($T_c <$ 9.2 K). Currents ranging from 0.4 mA to 0.8 mA are applied while the transverse voltage is measured as shown in figure 4b, maintaining similar current densities as in the bulk measurements.

The results of these experiments for a 6 µm thick sample are shown in Fig. 4c-d. First, we note that the $R_{xy}$ measurements resulting from a complete $\mu_0|H_{max}| = 1$ T loop protocol are vanishingly small ($H_{EB} < 5$ mT). Secondly, starting from ZFC conditions (green curve), we find that $H_c > 1$ T. This value should be compared with the $H_c \sim 0.2$ T that was previously obtained with an identical protocol on mm-sized samples. We find that the $H_c$ values are remarkably robust to radically different protocols. In particular, we did not observe demagnetization (Fig. 4c) or variation in $H_c$ under the application of $\mu_0 H_z = 6$ T (Fig. 4d). Moreover, we did not observe any domains during the magnetization reversal. The entire portion of the sample that was probed (more than 50 % of the area) changed magnetization abruptly in coincidence with the jump in $R_{xy}$ (Figure 4c insets, Supplementary movie 8). These large values of $H_c$ are similar to those previously reported in recent works[25–28].

Results from applying the same analysis to samples with different thicknesses (SM 4), revealed only a small, or no clear dependence of $H_c$ on the sample thickness. In addition, the demagnetization protocol has no effect on $H_c$ for any of the thicknesses tested. Finally, in all but the thickest (42 µm) sample we did not detect any measurable EB (Fig 4e-f).

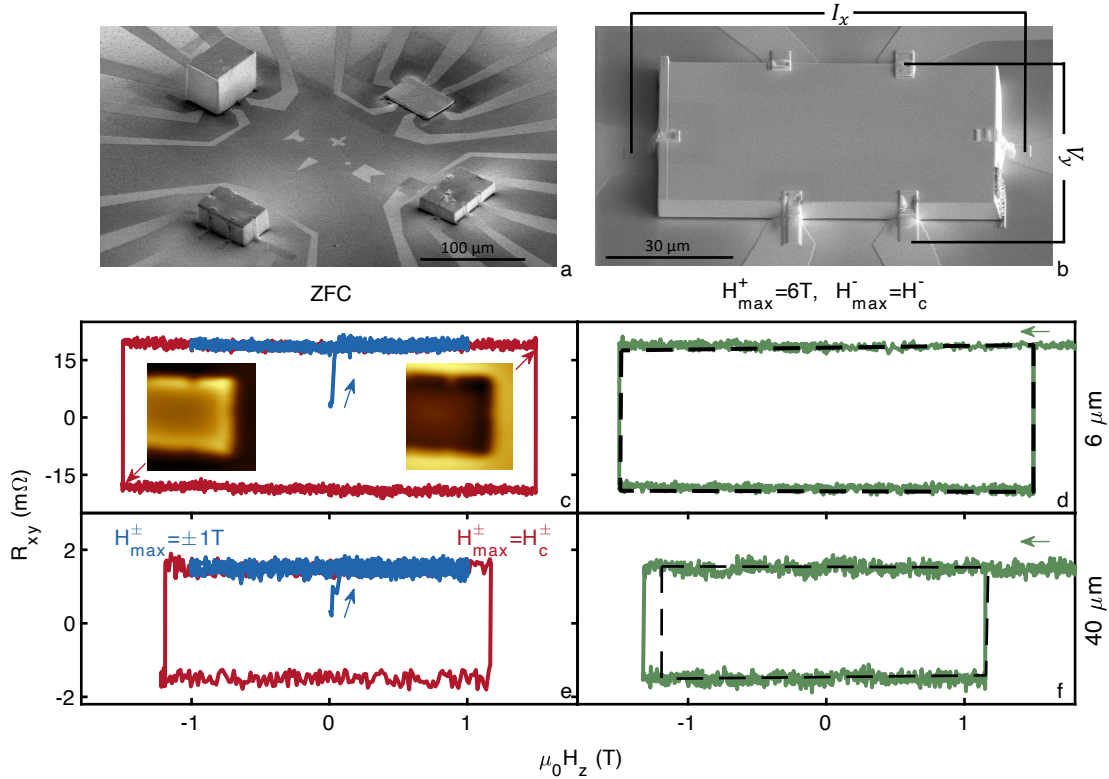

**Fig. 4 $R_{xy}$ measurements of Co$_3$Sn$_2$S$_2$ FIB samples with distinct thicknesses ranging from 6 to 42 μm at 10 K.** (**a**) SEM image of four different Co$_3$Sn$_2$S$_2$ single crystals cut by FIB with an area of 60 x 30 μm² and thicknesses ranging from 6 to 42 μm. (**b**) SEM image at higher magnification of the 6 μm thickness sample including a schematic of the measurement. **c-f** $R_{xy}$ measurements under different out-of-plane protocols. **c-d**. Sample thickness 6 μm. **e-f** Sample thickness 42 μm. (**c,e**) Blue: After ZFC, the field is swept three times between ±1T. Red: Finding the positive and negative coercive fields. Sweeping the field six times between the positive and negative coercive fields, does not change the coercive field, therefore no demagnetization process is possible in the FIB sample. Only two curves are shown for clarity. (**d,f**) Applying a positive 6 T field and then sweeping the field until the negative coercive field does not change the coercive field of the 6 μm sample. The negative coercive field of the 42 μm sample grown in 200 mT produces $H_{EB} = 100$ mT. The $R_{xy}$ measurements of samples with other thickness are presented in SM (4). C inset: SOT imaging of the 6 μm sample before and after $H_c^+$ presenting a single domine structure which flip at $H_c^+$. All images are 45x45 μm², pixel size 480 nm, acquisition time 9 min/image. The bright to dark color scale represents 10 mT and is the same for all images. See Supplementary Movie 8.

## Discussion

We now turn to discuss the physical consequences of our experimental findings. It is clear that $H_{EB}$ is **not** a static material property and can be tuned by applying IP and OOP field protocols. That $H_c^{\pm}$ evolves with $H_{max}^{\mp}$ well beyond the saturation field implies an undetectable evolution in the spin texture, which dramatically affects $H_c$ of the entire sample. The fact that the sign of $H_{EB}$ is determined by the direction of the IP field, even after that field is no longer applied, provides experimental evidence that IP spin canting determines the exchange bias of the OOP magnetic moment in both sign and size. Moreover, we show that these effects are not observable for microcrystals where, according to our magnetic imaging results, domain walls are not formed. This highlights the importance of domain walls and suggests they are linked to a hidden spin-texture.

Although the precise mechanism for the EB in this material remains to be identified, previous reports combined with our present findings, may point to a possible physical mechanism. A good candidate should be an uncompensated spin texture such as an AFM phase that would coexist with the dominating FM order. At the domain walls, ferromagnetism is weak and permits the appearance of competing magnetic orders. In that regard, Ref. [20] reported that the coexistence of AFM and FM survives all the way down to 25 K, in proximity to the saturation field where the presence of domain walls is expected. The AFM state presents a chiral magnetic structure of the IP spin component that could be caused by the inherent magnetic frustration of kagome lattice or Dzyaloshinsky-Moriya (DM) interlayer interaction. The two possible chiralities of the IP AFM state could be toggled by an IP field and affect $H_c$. Moreover, it is interesting to relate these results to other ferromagnets exhibiting exchange bias due to interlayer DM interaction[38] which are allowed by symmetry in $Co_3Sn_2S_2$. These results could also be related the memory effect emerging from the chiral domains in an in-plane AFM kagome[39].

Another tantalizing mechanism involves electronic boundary states that were theoretically predicted to form at the interface between two magnetic domains in magnetic Weyl semimetals. It was also predicted that Weyl nodes in this material should be sensitive to spin canting[33]. The presence of magnetic boundaries encompass exciting transport and electrostatic phenomena related to the axial magnetic field[19]. Though beyond the scope of the current work, our experimental system provides a unique opportunity to address these exotic phenomena present at magnetic domain walls.

**Methods**

**Single crystal growth.** Single crystals were grown from a stochiometric ratio of elements using the self-flux method (Sn flux). The elements were placed in $AlO_x$ crucible and sealed in an evacuated quartz tube.

**Transport measurements.** Transport measurements were performed at 4.2 K cooling inside a Liquid Helium dewar. A current of 2 mA was applied along the ab plane. Distance between current contacts was 1.6 mm and between $V_y$ contacts 1.2 mm. In all measurements, 25.4 μm Al wires were glued using silver epoxy to a 80 μm-thick sample as measured using a 3D profilometer.

**Scanning Squid-On-Tip microscopy:** The SOT was fabricated using self-aligned three-step thermal deposition of Pb at cryogenic temperatures, as described previously[36]. Supplementary Figure 1 shows the measured quantum interference pattern one of the SOTs used for this work with an effective diameter of 130 nm and a maximum critical current of 98 μA. The asymmetric structure of the SOT gives rise to a shift of the interference pattern resulting in good sensitivity in wide range of fields. All measurements were performed at 4.2 K in a low pressure He of ~1 mbar.

1.   Meiklejohn, W. H. & Bean, C. P. New Magnetic Anisotorpy. *Phys. Rev.* **105**, 904–913 (1957).

2.   Guang, Y. *et al.* Creating zero-field skyrmions in exchange-biased multilayers through X-ray illumination. *Nat. Commun.* **11**, 949 (2020).

3.   Ali, M. *et al.* Exchange bias using a spin glass. *Nat. Mater.* **6**, 70–75 (2007).

4.   Wang, B. M. *et al.* Large exchange bias after zero-field cooling from an unmagnetized state. *Phys. Rev. Lett.* **106**, 1–4 (2011).

5.   Canet, F., Mangin, S., Bellouard, C. & Piecuch, M. Positive exchange bias in ferromagnetic-ferrimagnetic bilayers: FeSn/FeGd. *Europhys. Lett.* **52**, 594–600 (2000).

6.   West, K. G. *et al.* Exchange bias in a single phase ferrimagnet. *J. Appl. Phys.* **107**, 113915 (2010).

7.   Nayak, A. K. *et al.* Design of compensated ferrimagnetic Heusler alloys for giant tunable exchange bias. *Nat. Mater.* **14**, 679–684 (2015).

8.   Migliorini, A. *et al.* Spontaneous exchange bias formation driven by a structural phase transition in the antiferromagnetic material. *Nat. Mater.* **17**, 28–34 (2018).

9.   Lachman, E. *et al.* Exchange biased anomalous Hall effect driven by frustration in a magnetic kagome lattice. *Nat. Commun.* **11**, 560 (2020).

10.  Hirschberger, M. *et al.* The chiral anomaly and thermopower of Weyl fermions in the half-Heusler GdPtBi. *Nat. Mater.* **15**, 1161–1165 (2016).

11.  Kuroda, K. *et al.* Evidence for magnetic weyl fermions in a correlated metal. *Nat. Mater.* **16**, 1090–1095 (2017).

12.  Liu, E. *et al.* Giant anomalous Hall effect in a ferromagnetic kagome-lattice semimetal. *Nat. Phys.* **14**, 1125–1131 (2018).

13.  Wang, Q. *et al.* Large intrinsic anomalous Hall effect in half-metallic ferromagnet Co3Sn2S2 with magnetic Weyl fermions. *Nat. Commun.* **9**, 3681 (2018).

14.  Sakai, A. *et al.* Giant anomalous Nernst effect and quantum-critical scaling in a ferromagnetic semimetal. *Nat. Phys.* **14**, 1119–1124 (2018).

15.  Ye, L. *et al.* Massive Dirac fermions in a ferromagnetic kagome metal. *Nature* **555**, 638–642 (2018).

16.  Morali, N. *et al.* Fermi-arc diversity on surface terminations of the magnetic Weyl semimetal Co3Sn2S2. *Science* **365**, 1286–1291 (2019).

17.  Yin, J.-X. *et al.* Negative flat band magnetism in a spin–orbit-coupled correlated kagome magnet. *Nat. Phys.* **15**, 443–448 (2019).

18.  Xu, Y. *et al.* Electronic correlations and flattened band in magnetic Weyl semimetal candidate Co3Sn2S2. *Nat. Commun.* **11**, 3985 (2020).

19.  Araki, Y. Magnetic Textures and Dynamics in Magnetic Weyl Semimetals. *Ann. Phys.* **532**, 1900287 (2020).

20.  Guguchia, Z. *et al.* Tunable anomalous Hall conductivity through volume-wise magnetic competition in a topological kagome magnet. *Nat. Commun.* **11**, 559 (2020).

21.  Liu, C. *et al.* Spin excitations and spin wave gap in the ferromagnetic Weyl semimetal Co3Sn2S2. *Sci. China Physics, Mech. Astron.* **64**, 217062 (2021).

22.  Liu, D. F. *et al.* Magnetic Weyl semimetal phase in a Kagomé crystal. *Science* **365**, 1282–1285 (2019).

23.  Schnelle, W. *et al.* Ferromagnetic ordering and half-metallic state of Sn2Co3S2. *Phys. Rev. B* **88**, 144404 (2013).

24. Shen, J. *et al.* On the anisotropies of magnetization and electronic transport of magnetic Weyl semimetal Co3Sn2S2. *Appl. Phys. Lett.* **115**, 212403 (2019).

25. Geishendorf, K. *et al.* Magnetoresistance and anomalous Hall effect in micro-ribbons of the magnetic Weyl semimetal Co3Sn2S2. *Appl. Phys. Lett.* **114**, 092403 (2019).

26. Geishendorf, K. *et al.* Signatures of the Magnetic Entropy in the Thermopower Signals in Nanoribbons of the Magnetic Weyl Semimetal Co3Sn2S2. *Nano Lett.* **20**, 300–305 (2020).

27. Tanaka, M. *et al.* Topological Kagome Magnet Co 3 Sn 2 S 2 Thin Flakes with High Electron Mobility and Large Anomalous Hall Effect . *Nano Lett.* **20**, 7476–7481 (2020).

28. Yang, S.-Y. *et al.* Field-Modulated Anomalous Hall Conductivity and Planar Hall Effect in Co 3 Sn 2 S 2 Nanoflakes. *Nano Lett.* **20**, 7860–7867 (2020).

29. Kassem, M. A., Tabata, Y., Waki, T. & Nakamura, H. Single crystal growth and characterization of kagomé-lattice shandites Co3Sn2−In S2. *J. Cryst. Growth* **426**, 208–213 (2015).

30. Kassem, M. A., Tabata, Y., Waki, T. & Nakamura, H. Structure and magnetic properties of flux grown single crystals of Co3−Fe Sn2S2 shandites. *J. Solid State Chem.* **233**, 8–13 (2016).

31. Windsor, Y. W., Gerber, A. & Karpovski, M. Dynamics of successive minor hysteresis loops. *Phys. Rev. B* **85**, 064409 (2012).

32. Windsor, Y. W., Gerber, A., Korenblit, I. Y. & Karpovski, M. Time dependence of magnetization reversal when beginning with pre-existing nucleation sites. *J. Appl. Phys.* **113**, 223902 (2013).

33. Ghimire, M. P. *et al.* Creating Weyl nodes and controlling their energy by magnetization rotation. *Phys. Rev. Res.* **1**, 032044 (2019).

34. Uri, A. *et al.* Nanoscale imaging of equilibrium quantum Hall edge currents and of the magnetic monopole response in graphene. *Nat. Phys.* **16**, 164–170 (2020).

35. Uri, A. *et al.* Mapping the twist-angle disorder and Landau levels in magic-angle graphene. *Nature* **581**, 47–52 (2020).

36. Vasyukov, D. *et al.* A scanning superconducting quantum interference device with single electron spin sensitivity. *Nat. Nanotechnol.* **8**, 639–644 (2013).

37. Anahory, Y. *et al.* SQUID-on-tip with single-electron spin sensitivity for high-field and ultra-low temperature nanomagnetic imaging. *Nanoscale* **12**, 3174–3182 (2020).

38. Fernández-Pacheco, A. *et al.* Symmetry-breaking interlayer Dzyaloshinskii–Moriya interactions in synthetic antiferromagnets. *Nat. Mater.* **18**, 679–684 (2019).

39. Li, X. *et al.* Chiral domain walls of Mn3Sn and their memory. *Nat. Commun.* **10**, 3021 (2019).

Acknowledgements

We would like to thank B. Yan, Z. Ovadyahu, A. Vaknin, and A. Capua for fruitful discussions. We would like to thank S. Ben Atar for manufacturing the SOT microscope, A. Vakahi for fabricating the FIB samples, N. Katz for growing the Nb thin films and to E. Sabag which was responsible for the construction of the Quantum Imaging Lab. This work was supported by the European Research Council (ERC) Foundation grant No. 802952 and the Israel Science Foundation (ISF) grant No. 649/17 and 2178/17. The international collaboration on this work was fostered by the EU-COST Action CA16218. S. Gazit acknowledges support from the Israel Science Foundation, Grant No. 1686/18. J.G.Analytis and E. Lachman acknowledges support from the Gordon and Betty Moore foundation's EPiQS Initiative through Grant GBMF9067. E. Lachman is an Awardee of the Weizmann Institute of Science - National Postdoctoral Award Program for Advancing Women in Science. F. Toric is an awardee of the Hebrew University Center for Nanoscience and Nanotechnology Postdoctoral Fellowship.

**Supplementary Figure 1: SOT characterization**

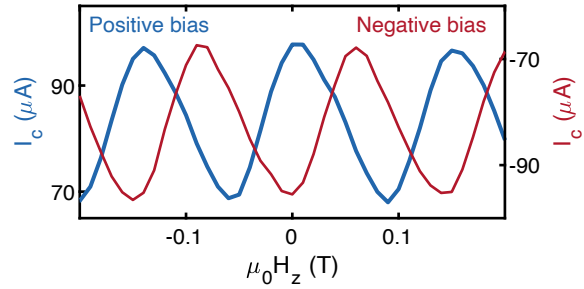

**Supplementary Figure 1. Quantum interference pattern of the SQUID-on-tip (SOT).** The critical current $I_c$ of one of the SOT used in this work as a function of the applied out-of-Plane field $H_z$. Blue: Positive bias, red: Negative bias. The period of 150 mT of the quantum interference corresponds to an effective diameter of 130 nm of the SOT.

**Supplementary Figure 2: Controlling the exchange bias**

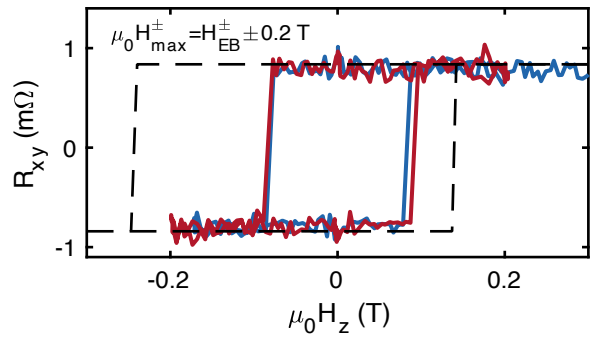

**Supplementary Figure 2. Controlling the exchange bias AHE of Co$_3$Sn$_2$S$_2$ single crystal sample using different out-of-plane protocols at 4.2 K.** $R_{xy}$ as a function of $H_z$. Eliminated $H_{EB}$ within our experimental resolution and decreasing the coercive to minimum value of $H_c^\pm < 0.1T$, by applying a non-symmetric loop of $\mu_0 H_{max}^\pm = H_{EB} \pm 200$ mT. Black dashed curve: initialized the sample by applying once a loop with $\mu_0|H_{max}^\pm| = 1$ T.

**Supplementary Figure 3: In-Plane measurements**

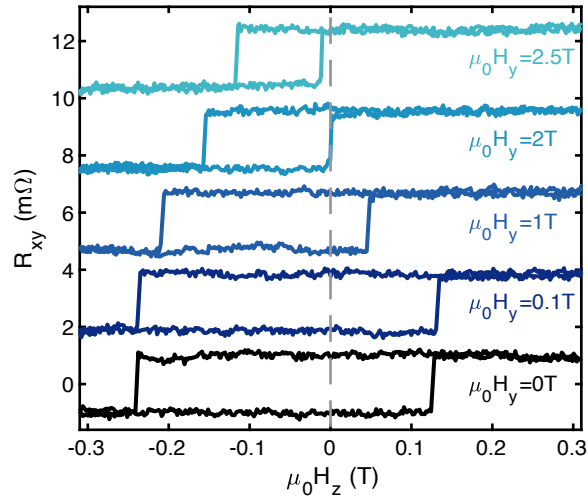

**Supplementary Figure 3. Hall resistance measurements of $Co_3Sn_2S_2$ as a function of in-plane field at 5 K.** Hall resistance measurements for a $\mu_0|H_{max}^{\pm}| = 1$ T protocol while applying a an in-plane field of $H_y = 0, 0.1, 1, 2, 2.5$ T corresponding to $H_{EB} = 56, 52, 78, 78, 64$ mT, respectively.

**Supplementary Fig 4: FIB transport 1 and 20 µm thick**

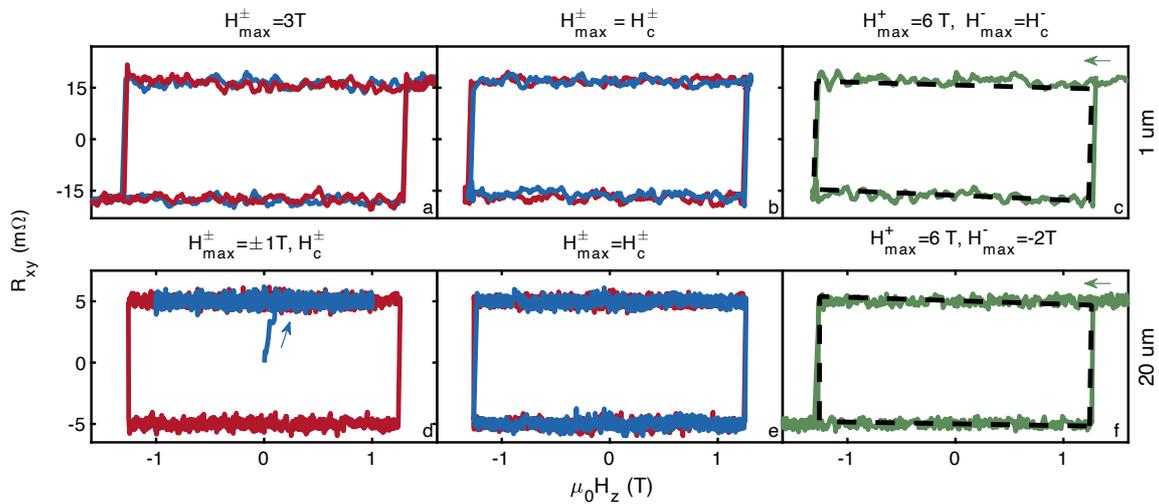

**Supplementary Figure 4.** $R_{xy}$ **measurements of $Co_3Sn_2S_2$ FIB samples in thickness of 1 and 20 µm at 10 K. a-f** $R_{xy}$ measurements under different out-of-plane protocols of $Co_3Sn_2S_2$ single crystal cut by FIB in area of 60 x 30 µm² in various thicknesses. **a-c.** Sample thickness 1 µm. **d-f** Sample thickness 20 µm. **(a)** A ±3 T protocol applied resolving no change in $H_c$ and zero EB. **(b)** The field swept between positive and negative coercive field six times resulting no change in coercive field, only two curves shown for clarity. **(c)** A positive 6 T field was applied and then the field swept till the negative coercive field. No change in the $H_c$ obtained. **(d)** Blue: After ZFC the field swept between ±1T three times. Red: Finding the positive and negative coercive field. No EB was measured. **(e)** Same as **b.** **(f)** A positive 6 T field was applied and then the field swept to -2 T. No change in $H_c$ and zero EB.

**Supplementary Figure 5: Sample characterization**

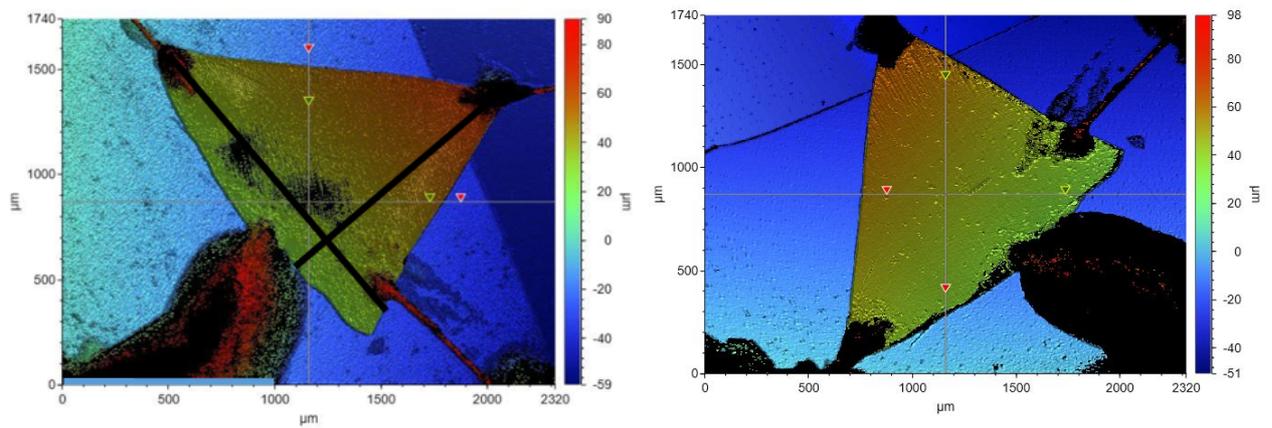

**Supplementary Figure 5. Sample dimensions and tilt.** Image of the sample using a 3D interferometer.

**Supplementary Figure 6: EDS measurements of the FIB sample**

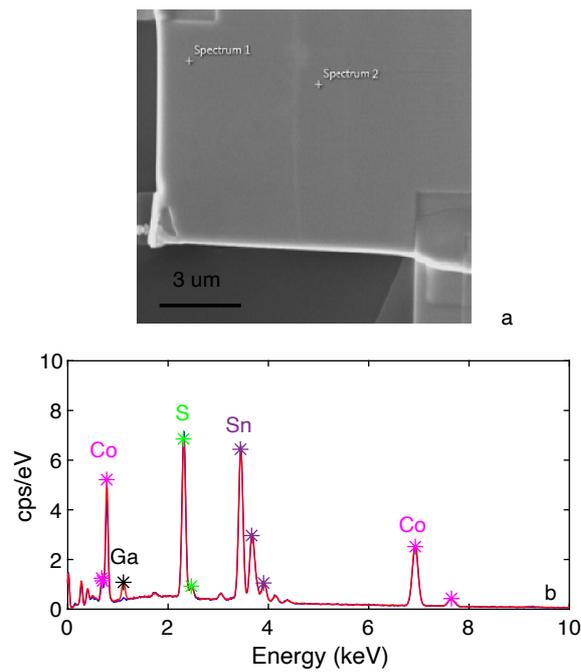

**Supplementary Figure 6. Energy dispersive spectroscopy (EDS) measurements of the purity of the Co3Sn2S2 FIB sample. a.** A SEM image of FIB sample with two arbitrary chosen positions indicates on the EDS measurement. **b**. Blue: Energy spectrum of area 1, Red: Energy spectrum of area 2. The elements symbols label above each peak.

**Supplementary Note 1: SOT characterization**

Supplementary figure 1 shows the quantum interference pattern typical for a SQUID. The SOT response to magnetic field is linked to the derivative of that pattern. To get good images, the SOT has to be field biased in a region where the interference pattern is linear, and one must avoid the regions where the response is zero (blind spots). The field period of this pattern is determined by the SQUID loop diameter so that each period represents a magnetic flux equal to the quantum of flux $\phi_0 = h/2e \approx 20.67$ Gµm². This means that a smaller SQUID loop will yield a large field period with larger linear regions but with larger blind spots. Depending on the experimental requirements, optimal SQUID loop size is chosen. To mitigate the effect of blind spots, we use the fact that our SOTs are often composed of asymmetric junctions. This makes the interference pattern to have a different field offset for a different direction of the current running through the SQUID (Supplementary Figure 1, blue and red curves).

The magnetic contrast in $Co_3Sn_2S_2$ is on the order of tens of mT, which imposes that the interference pattern must be linear over that range of field. That implies a SQUID loop with a diameter below 150 nm. However, having a SQUID with a loop too small, of 50 nm for example, would yield a blind spot of +/- 0.4 T around zero field. That would be problematic to get good images in the ZFC state. For this reason, all the SOTs used in this work have a SQUID loop of about 110-130 nm.

**Supplementary Note 2: Controlling the exchange bias**

In addition to the protocols presented in fig. 1 here, we present another protocol which controls the coercive field $H_c$ and therefore the exchange bias $H_{EB}$. We now demonstrate that it is possible to reduce $H_{EB}$ by applying a non-symmetric loop with $\mu_0|H_{max}^{\pm}| < 1$ T. The results are shown in Supplementary Fig 2. First, we initialized the sample by applying once a loop with $\mu_0|H_{max}^{\pm}|$ = 1 T (Black curve). We obtain the expected curve shown in fig. 1b. Then we sweep the field up to $\mu_0 H_{max}^{\pm} = H_{EB} \pm 200$ mT, rather than sweeping symmetrically around zero. When sweeping the field from $-1$ T towards the positive direction, $R_{xy}$ jumps at 143 mT. The field is then ramped up to $\mu_0 H_{max}^{\pm} = H_c^+ + 200$ mT $= 343$ mT (Blue curve) before ramping the field in the negative direction since $H_c^-$ is not well defined before the first loop is completed and $H_{EB}$ cannot be calculated. $H_c^-$ is then reduced to $-86$ mT so the field is swept down to $-171$ mT ($\mu_0 H_{max}^{-} = H_{EB} - 200 = -171$ mT). Ramping the field in the other direction, we found $H_c^+ = 88$ mT, meaning we have eliminated the EB within our experimental resolution. Red: More loops of $\mu_0 H_{max}^{\pm} = H_{EB} \pm 200$ were performed resolving similar values of $H_c$ to confirm the reproducibility (only one shown for clarity).

**Supplementary Note 3: In-Plane measurements**

To verify the influence of the IP component on $H_c^{\pm}$, we measure $R_{xy}(H_z)$ for different $H_y$. In each measurement, we applied the $\mu_0|H_{max}^{\pm}|$ = 1 T protocol. In Supplementary Figure 3 we present a sequence of loops in $H_z$ in presence of different $H_y$ ranging from 0 to 2.5 T resulting in a slight variation in $H_{EB}$, starting from 56 mT and reaching a maximum at $\mu_0 H_y = 1$ and 2 T and reducing to 64 mT for $\mu_0 H_y = 2.5$ T. Some of these variations could easily be explained by a slight tilt in the sample, which is evaluated to be ~1-2° which causes the IP field to generate a small OOP field. However, a clear decrease of both $|H_c^+|$ and $|H_c^-|$ is observed. This cannot be explained by a sample misalignment and shows the influence of the presence of an IP field on $H_c^{\pm}$.

**Supplementary Note 4: FIB transport 1 and 20 μm thick**

In the main text, we present two samples with 6 and 40 μm thickness. Here we present two more samples with 1 and 20 μm. For the smallest sample achievable by FIB, which was 1 μm thick, $H_c = 1.3$T remained constant for all protocols and no measurable EB was observed. In Supplementary Fig. 4a we present the $\mu_0|H_{max}| = 3$ T protocol. In Fig. 4b we present the $\mu_0|H_{max}| = H_c^{\pm}$ protocol, and in Fig. 4c, we present an asymmetric protocol of $H_{max}^+ = 6$ T and $H_{max}^- = H_c^-$, no sizeable changes in $H_c$ were found in none of the protocols.

Then we present the measurement of the 20 μm thick sample with slightly different protocols. In Fig. 4d we present the $\mu_0|H_{max}| = 1$ T protocol from the ZFC state to saturated state (blue curve). As observed in other FIB samples, after performing the loop with $\mu_0|H_{max}| = 1$ T, the coercive field was not reached. Increasing $|H_{max}|$ we found that $H_C > 1$ T, which is in the same order as other FIB samples. Moreover, no measurable EB was observed. In Fig. 4e we applied the $\mu_0|H_{max}| = H_c^{\pm}$ protocol, no significant change in $H_c$ was observed even after 6 loops (only two shown for clarity). In Fig. 4f we present an asymmetric protocol of $H_{max}^+ = 6$ T and $H_{max}^- = -2$ T, no significant changes in $H_c$ were found.

**Supplementary Note 5: Sample characterization**

The sample measured in a 3D optical profilometer in order to measure the Co$_3$Sn$_2$S$_2$ crystal dimensions and to estimate it's tilt. The thickness of the crystal is ~ 80 μm, the distance between current contacts was 1.6 mm and between V$_y$ contacts 1.2 mm, resulting in lateral dimensions of ~ 0.95 mm$^2$ and a volume of ~0.075 mm$^3$. The tilt of the sample measured in the direction of the applied in-plane (IP) field estimate to be -1.2° and in the order of 1° in the transverse direction. Therefore, we conclude that a small correction of the IP field transport measurement is required.

**Supplementary Note 6: EDS measurements of the FIB sample**

We used energy-dispersive spectroscopy (EDS) to perform an elemental analysis of the sample. Few arbitrary positions were chosen, resolving the expected ratio of Co$_3$Sn$_2$S$_2$ as reported below. In Supplementary Fig. 6a we present a SEM image of the FIB sample, with labels Spectrum 1 and Spectrum 2 indicating 2 regions over which the spectrum was collected, and the corresponding spectra are plotted in Supplementary Fig 6b Spectrum 1 in blue, and spectrum 2 in red. The spectra are identical and overlap. The peaks can be assigned to all expected chemical elements: cobalt, tin, and sulfur. A small peak at 1.1 keV is associated with gallium (Ga) contamination. The gallium peak comes from the ion beam of the FIB and is expected to be concentrated at the surface, roughly less than a few hundred nanometers. The results of the atomic percentage measurements are Co=42.7%, Sn=29.7%, and S=27.6%, which agrees with the ratio of a pure stoichiometric crystal. Therefore, the EDS measurements rule out the possibility that the difference between FIB and bulk are influenced by contamination of other elements.

**Movies:**

**Supplementary Movie 1:**

Movie of magnetic domains evolution imaged with the SOT from Zero Field Cooling (ZFC) to the saturated state in $Co_3Sn_2S_2$. After ZFC, $H_z$ was ramped from 25 to 100 mT. Magnetic features of a few microns exhibiting a large magnetic contrast (~ 50 mT) were observed. By increasing $H_z$, the domains that are anti-aligned with the field shrink while the ones that are parallel growing. From $\mu_0 H_z = 95$ mT, the magnetic contrast drops below 1.5 mT and the $B_z(x,y)$ stops evolving on that scale. The frame size was $45 \times 45$ µm$^2$; pixel size was $480 \times 480$ nm$^2$, acquisition time was 24 min/frame and $H_z$ was increased in 5 mT step. All frames are with the same color scale of 50 mT; when high magnetic contrast appears, the color scale is intentionally saturated at the edge for clarity. Selected frames from the movie are shown Fig. 3c-f of the main text.

**Supplementary Movie 2:**

Movie of magnetic domains evolution imaged with the SOT from the demagnetized state to the saturated state in $Co_3Sn_2S_2$. After applying the demagnetization protocol described in the main text, $H_z$ was ramped from 30 to 100 mT. $B_z(x,y)$ exhibits large stripes of roughly 10 µm width and more than 45 µm length yielding a signal of 50 mT. These features are significantly different from the ones observed in the ZFC (Fig 4c-f and movie 1) while $R_{xy}$ is comparable. On the other hand, these feature show similarities with the transitory state (Fig 4k-n and movie 3), although here $R_{xy}$ evolves from vanishingly small value to the saturated value as shown in the panel below. From $\mu_0 H_z = 95$ mT, the magnetic contrast drops below 1.5 mT and the $B_z(x,y)$ stops evolving on that scale. The frame size was $45 \times 45$ µm$^2$; the pixel size was $980 \times 980$ nm$^2$, the acquisition time was 8 min/frame and $H_z$ was increased in 5 mT step. All frames are with a uniform color scale of 50 mT, when high magnetic contrast appears, the color scale is intentionally saturated at the edge for clarity. Selected frames from the movie are shown Fig. 3g-j.

**Supplementary Movie 3:**

Movie of magnetic domains evolution imaged with the SOT from one saturated state to the other saturated state through an intermediate transitory phase in $Co_3Sn_2S_2$. After applying the $|H_{max}^{\pm}| = \pm 200$ mT protocol described in the main text, $H_z$ was ramped from 10 to 100 mT. $B_z(x,y)$ exhibits large stripes of roughly 10 µm width and more than 45 µm length yielding a signal of 50 mT. These features are significantly different from the ones observed in the ZFC (Fig 4c-f and movie 1). On the other hand, these feature show similarities with the demagnetized state (Fig 4g-j and movie 2), although here $R_{xy}$ evolves sharply at the coercive field and reaches the saturated value in 15 mT as shown in the panel below. From $\mu_0 H_z = 95$ mT, the magnetic contrast drops below 1.5 mT and the $B_z(x,y)$ stops evolving on that scale. The frame size was $45 \times 45$ µm$^2$, the pixel size was $980 \times 980$ nm$^2$, the acquisition time was 8 min/frame and $H_z$ was increased in 2 mT step. All frames are with a constant color scale of 50 mT; when high magnetic contrast appears, the color scale is intentionally saturated at the edge for clarity. Selected frames from the movie are shown Fig. 3k-n.

**Supplementary Movie 4:**

Movie of magnetic domains evolution imaged with the SOT from Zero Field Cooling (ZFC) to the saturated state in $Co_3Sn_2S_2$. After ZFC, $H_z$ was ramped from -20 to -100 mT. Magnetic features of a few micron exhibiting a large magnetic contrast (~ 20 mT) were observed. By increasing $H_z$, the domains that are anti-aligned with the field shrink while the ones that are parallel growing. From $\mu_0 H_z = -95$ mT, the magnetic contrast drops below 1.5 mT and the $B_z(x,y)$ stops evolving on that scale. The frame size was $45 \times 45$ µm$^2$, pixel size was $480 \times 480$ nm$^2$, acquisition time was 24 min/frame and $H_z$ was increased in 1.5 mT step. All frames are with the same color scale of 20 mT; when high magnetic contrast appears, the color scale is intentionally saturated at the edge for clarity.

**Supplementary Movie 5:**

Movie of magnetic domains evolution imaged with the SOT from Zero Field Cooling (ZFC) to the saturated state in $Co_3Sn_2S_2$. After ZFC, $H_z$ was ramped from 20 to 100 mT. Magnetic features of a few microns exhibiting a large magnetic contrast ($\sim 60$ mT) were observed. By increasing $H_z$, the domains that are anti-aligned with the field shrink while the ones that are parallel growing. From $\mu_0 H_z = 95$ mT, the magnetic contrast drops below 1.5 mT and the $B_z(x,y)$ stops evolving on that scale. The frame size was $45 \times 45$ μm², pixel size was $480 \times 480$ nm², acquisition time was 24 min/frame and $H_z$ was increased in 1.5 mT step. All frames are with the same color scale of 60 mT; when high magnetic contrast appears, the color scale is intentionally saturated at the edge for clarity.

**Supplementary Movie 6:**

Movie of magnetic domains evolution imaged with the SOT from the demagnetized state to the saturated state in $Co_3Sn_2S_2$. After applying the demagnetization protocol described in the main text, $H_z$ was ramped from -30 to -100 mT. $B_z(x,y)$ exhibits large stripes of roughly 10 μm width and more than 45 μm length yielding a signal of 30 mT. These features are significantly different from the ones observed in the ZFC (Fig 4c-f and movie 1) while $R_{xy}$ is comparable. On the other hand, these feature show similarities with the transitory state (Fig 4k-n and movie 3), although here $R_{xy}$ evolves from vanishingly small value to the saturated value as shown in the panel below. From $\mu_0 H_z = -95$ mT, the magnetic contrast drops below 1.5 mT and the $B_z(x,y)$ stops evolving on that scale. The frame size was $45 \times 45$ μm²; the pixel size was $1.5 \times 1.5$ μm², the acquisition time was 8 min/frame and $H_z$ was increased in 10 mT step. All frames are with a constant color scale of 30 mT; when high magnetic contrast appears, the color scale is intentionally saturated at the edge for clarity.

**Supplementary Movie 7:**

Movie of magnetic domains evolution imaged with the SOT from one saturated state to the other saturated state through an intermediate transitory phase in $Co_3Sn_2S_2$. After applying the $|H_{max}^{\pm}| = \pm 200$ mT protocol described in the main text, $H_z$ was ramped from -30 to -100 mT. $B_z(x,y)$ exhibits large stripes of roughly 10 μm width and more than 45 μm length yielding a signal of 50 mT. These features are significantly different from the ones observed in the ZFC (Fig 4c-f and movie 1). On the other hand, these feature show similarities with the demagnetized state (Fig 4g-j and movie 2), although here $R_{xy}$ evolves sharply at the coercive field and reaches the saturated value in 15 mT as shown in the panel below. From $\mu_0 H_z = -95$ mT, the magnetic contrast drops below 1.5 mT and the $B_z(x,y)$ stops evolving on that scale. The frame size was $45 \times 45$ μm², the pixel size was $1.5 \times 1.5$ μm², the acquisition time was 8 min/frame and $H_z$ was increased in 5 mT step. All frames are with a constant color scale of 50 mT; when high magnetic contrast appears, the color scale is intentionally saturated at the edge for clarity.

**Supplementary Movie 8:**

Movie of magnetic evolution of a 6 μm thick $Co_3Sn_2S_2$ sample imaged with the SOT from one saturated state to the other saturated state. After applying the $|H_{max}^{\pm}| = \pm 1.7$ T protocol $H_z$ was ramped from 1.6 to 1.7 T. At the coercive field $H_c^+$ the polarization flip at once accompanied by a jump of the $R_{xy}$ as shown in the panel below. We did not observe any magnetic domains during the magnetization reversal. These images are equivalent to the saturated state images of the bulk (Fig 4j,n). The SOT is not sensitive above $\sim 1$ T since that field exceeds the critical field of the superconducting Pb at 4.2 K in that geometry. For that reason, the field was reduced to XX T before acquiring each frame. The frame size was $45 \times 45$ μm²; the pixel size was $480 \times 480$ μm², the acquisition time was 9 min/frame and $H_z$ was increased in 5 mT step. All frames are with a uniform color scale of 10 mT. Selected frames from the movie are shown in Fig. 4c inset.